\documentstyle[amssymb,12pt]{article}

\begin{document}

\title{Accelerated Levi-Civita-Bertotti-Robinson Metric in $D$-Dimensions}
\author{ Metin G{\" u}rses\\
{\small Department of Mathematics, Faculty of Sciences}\\
{\small Bilkent University, 06800 Ankara-Turkey}\\
{\small email: gurses@fen.bilkent.edu.tr}\\
\\
{\" O}zg{\" u}r Sar{\i}o\u{g}lu  \\
{\small Department of Physics, Faculty of Arts and  Sciences}\\
{\small Middle East Technical University, 06531 Ankara-Turkey}\\
{\small email: sarioglu@metu.edu.tr}}
\begin{titlepage}
\maketitle
\begin{abstract}
A conformally flat accelerated charge metric is found in an
arbitrary dimension $D$. It is a solution of the
Einstein-Maxwell-null fluid equations with a cosmological constant
in $D \ge 4$ dimensions. When the acceleration is zero our
solution reduces to the Levi-Civita-Bertotti-Robinson metric. We
show that the charge loses its energy, for all dimensions, due to
the acceleration.

\end{abstract}


\end{titlepage}

\section*{I. Introduction}

Levi-Civita-Bertotti-Robinson (LBR) metric is one of the classical
metrics in general relativity. It is the only Einstein-Maxwell
solution which is homogeneous and has homogeneous non-null Maxwell
field \cite{lev}-\cite{kram}. LBR space-time is a product of two
spaces of constant curvature, namely it is $AdS_{2} \times S^2$.
Due to this  property LBR metric arises also in supergravity
theories \cite{gib}-\cite{kal}. LBR type of space-times show up
also as the space-time regions closer to the horizons of extreme
Reissner-Nordstrom black hole geometries \cite{gur4}, \cite{cor}.
Its curvature invariants are all constant and the electromagnetic
field tensor is covariantly constant. This property leads to the
result that LBR metric is an exact solution of any theory of
gravitation coupled with a $U(1)$ gauge field \cite{gur4},
\cite{gur5}.

In this work we generalize the LBR metric in $D$-dimensions by
introducing acceleration. A similar generalization was done long
ago for the Reissner-Nordstrom metric in four dimensions
\cite{bv1}. Recently we extended the Bonnor-Vaidya formalism to
Reissner-Nordstrom metric in $D$-dimensions \cite{gur1},
\cite{gur2}. We used an arbitrary curve $C$ in $D$-dimensional
Minkowski space-time $M_{D}$ and examined in detail the
Einstein-Maxwell-null  dust \cite{gur1} and
Einstein-Born-Infeld-null dust field equations \cite{gur2},
Yang-Mills equations \cite{ozg}, and Li{\' e}nard-Wiechert
potentials in even dimensions \cite{gur3}. In the first three
works \cite{gur1}, \cite{gur2}, \cite{ozg}, we found some new
solutions generalizing the Tangherlini \cite{tan}, Pleba{\' n}ski
\cite{pleb}, and Trautman \cite{trt} solutions, respectively. The
last one, \cite{gur3}, indicates that the accelerated scalar or
vector charged particles in even dimensions lose energy.

 Our conventions are
similar to the conventions of our earlier works  \cite{gur1},
\cite{gur2}, \cite{gur3}. In a $D$-dimensional Minkowski
space-time $M_{D}$, we  use a parametrized curve $C= \{x^{\mu} \in
M_{D}|\,\, x^{\mu}=z^{\mu}(\tau)\,, \mu=0,1,2, \cdots, D-1 \, ,
\tau \in I \}$ such that $\tau$ is a parameter of the curve and
$I$ is an interval on the real line ${\mathbb R}$. We define the
world function $\Omega$ as

\begin{equation}
\Omega=\eta_{\mu \nu}\,(x^{\mu}-z^{\mu}(\tau))\,(x^{\nu}-z^{\nu}(\tau)),
\label{dist}
\end{equation}

\noindent where $x^{\mu}$  is a point not on the curve $C$. There
exists a point $z^{\mu}(\tau_{0})$ on the non-space-like curve $C$
which is also on the light cone with the vertex located at the
point $x^{\mu}$, so that $\Omega(\tau_{0})=0$. Here $\tau_{0}$ is
the retarded time. By using this property we find that

\begin{equation}\label{lam1}
\lambda_{\mu} \equiv \partial_{\mu}\, \tau_{0} =
{x^{\mu}-z^{\mu}(\tau_{0}) \over R},
\end{equation}

\noindent where $R \equiv
\dot{z}^{\mu}(\tau_{0})\,(x_{\mu}-z_{\mu}(\tau_{0}))$ is the
retarded distance. Here a dot over a letter denotes
differentiation with respect to $\tau_{0}$. It is easy to show
that $\lambda_{\mu}$ is null and satisfies

\begin{eqnarray}
\lambda_{\mu, \nu}&=&{1 \over R}\, [\eta_{\mu \nu}-\dot{z}_{\mu}\,
\lambda_{\nu}-\dot{z}_{\nu}\, \lambda_{\mu}-(A-\epsilon)\,
\lambda_{\mu}\, \lambda_{\nu}],\\
R_{,\mu}&=&\dot{z}_{\mu}+(A -\epsilon ) \lambda_{\mu},
\end{eqnarray}

\noindent where $A \equiv \ddot{z}^{\mu}\,
(x_{\mu}-z_{\mu}(\tau_{0}))$ and $\dot{z}^{\mu}\,
\dot{z}_{\mu}=\epsilon= -1,0$. Here $\epsilon=-1$ and $\epsilon=0$
correspond to the time-like and null velocity vectors,
respectively. In this work we shall consider only the case where
the velocity vector is time-like ($\epsilon=-1$). One can also
show explicitly that $\lambda^{\mu}\, \dot{z}_{\mu}=1$ and
$\lambda^{\mu}\, R_{,\, \mu}=1$. If one defines $a \equiv {A /
R}=\lambda^{\mu} \ddot{z}_{\mu}$, then

\begin{equation}
\lambda^{\mu}\, a_{, \, \mu}=0.
\end{equation}

\noindent Furthermore defining (letting $a_{0} \equiv a$)

\begin{equation}
a_{k} \equiv \lambda_{\mu}\, {d^{k+2} \, z^{\mu}(\tau_{0}) \over d
\tau_{0} ^{k+2}},~~~k=0,1,2, \cdots \label{aks}
\end{equation}

\noindent one can show that

\begin{equation}
 \lambda^{\mu}\, a_{k,\, \mu}=0,~~~~ \forall k=0,1,2, \cdots
\end{equation}

In the next section, using \cite{gur4},  we present the LBR metric
in  arbitrary $D$-dimensions.  In Section III we give the
accelerated LBR metrics in $D$-dimensions. Here we find the rate
of energy loss for all $D$. In Section IV we consider a special
curve $C$ where the null fluid disappears. In particular, taking a
curve $C$ which corresponds to a constant acceleration, we find a
solution of the Einstein-Maxwell field equations in
$D$-dimensions. We show that this is also the LBR metric on an
accelerated reference frame.

\section*{II. Levi-Civita-Bertotti-Robinson Metrics}

We assume a spherically symmetric, static space-time in
$D$-dimensions. Among such a class of space-times, LBR geometry is
a product space-time $AdS_{2} \times S^{D-2}$.  It  has been
studied for several purposes \cite{gur4}, \cite{gur5}, (see also
\cite{cor}). We have the following theorem.

\vspace{0.3cm}

\noindent {\bf Theorem 1}.\, {\it Let  $g_{\mu \nu}$ be the metric
and $F_{\mu \nu}$ be the Maxwell field   given by
\begin{eqnarray}
g_{\mu \nu}&=&{q^2 \over r^2}\,[-t_{\mu}\,t_{\nu}+c_{0}^2\,
k_{\mu}\, k_{\nu}+r^2\, h_{\mu \nu}],\\
F_{\mu \nu}&=&{c_{0} \over r}\,(t_{\mu}\, k_{\nu}-t_{\nu}\,
k_{\mu}),
\end{eqnarray}
where $h_{\mu \nu}$ is the metric of the $(D-2)$-dimensional
sphere $S^{D-2}$, $c_{0}$, $q$ are constants,
$t_{\mu}=\delta_{\mu}^{0}$, $k_{\mu}=\delta_{\mu}^{r}$. Then they
solve the Einstein-Maxwell field equations with a cosmological
constant
\begin{equation}
G_{\mu \nu}={Q^2 \over c_{0}^2}\, [F_{\mu}\,^{\alpha}\,F_{\nu
\alpha}-{1 \over 4} (F^{\alpha \beta}\,F_{\alpha \beta}) g_{\mu
\nu}]+\Lambda g_{\mu \nu},
\end{equation}
where
\begin{equation}
Q^2=q^2[(D-3)\,c_{0}^2+1],~~~\Lambda={1 \over 2q^2}\,[{1 \over
c_{0}^2}-(D-3)^2].
\end{equation}}

\vspace{0.3cm}

\noindent Here $Q$ is the electric charge. This metric describes
the near horizon region of the charged Tangherlini metric with a
cosmological constant \cite{gur4}, \cite{cor} and moreover  is
conformally flat if $c_{0}=1$. In this case the cosmological
constant vanishes only when $D =4$. All the curvature invariants
are constants, being functions of the constants, $c_{0}, Q$ and
$D$. These metrics describe the geometry of black holes in the
neighborhood of their outer horizons \cite{gur4}, \cite{gur5},
\cite{cor}. Our purpose in this work is to generalize this
solution when the charge moves on a curve $C$ described in the
introduction.

\section*{III. Accelerated Bertotti-Robinson Metric in $D$-Dimensions}

In this section, we shall generalize the LBR metric by introducing
acceleration using the curve kinematics given in the introduction.
We shall consider the case when the space-time is conformally
flat. We assume that the metric and the electromagnetic vector
potential are given by

\begin{equation}
g_{\mu \nu}=e^{\psi}\, \eta_{\mu \nu},~~~ A_{\mu }=S \,
\dot{z}_{\mu},
\end{equation}

\noindent where $\eta_{\mu \nu}$ is the Minkowski metric, $\psi$
and $S$ are functions of $R$ only. Then we have

\vspace{0.3cm}

\noindent {\bf Theorem 2}.\, {\it In an arbitrary dimension $D$
\begin{eqnarray}
ds^2&=&{R_{0}^2 \over R^2}\, \,\eta_{\mu \nu}\,dx^{\mu}\,dx^{\nu},\\
A_{\mu}&=&e\,{\dot{z}_{\mu} \over R},
\end{eqnarray}
are the solutions of the Einstein-Maxwell-null (pressureless)
fluid equations with cosmological constant $\Lambda$,

\begin{equation}
G_{\mu \nu}=\kappa\, [F_{\mu}\,^{\alpha}\, F_{\nu \alpha}-{1 \over
4}\, (F^{\alpha \beta}\, F_{\alpha \beta})\,\, g_{\mu \nu}]+
\kappa\, \rho \lambda_{\mu}\, \lambda_{\nu}+\Lambda g_{\mu \nu}.
\end {equation}
\noindent The energy density of the fluid and the cosmological
constant are given by
\begin{eqnarray}
\kappa\,\rho&=&(D-2)\, (a_{1}-\ddot{z}^{\alpha}\, \ddot{z}_{\alpha}),\\
\Lambda&=&-{(D-2)(D-4) \over 2\,R_{0}^2},~~e^2={D-2 \over \kappa}
R_{0}^2,
\end{eqnarray}
where $R_{0}$ is a constant, $R$ is the retarded distance
described  in the introduction, $\lambda_{\mu}$ is the null vector
defined in (\ref{lam1}) and $a_{1}$ is defined in (\ref{aks}).
Furthermore, the current vector $J^{\mu}=\, \nabla_{\nu}\, F^{\mu
\nu}$ vanishes for all $D$ (except on the curve $C$, see Remark 1
below).}

 \vspace{0.3cm}

\noindent First of all when the curve $C$ is a straight line, the
solution given in Theorem 2 reduces to the one in Theorem 1 with
$c_{0}=1$. Moreover we observe that the metric and the
electromagnetic fields for all dimensions  have the same form.
This is interesting because in the case of accelerated charges in
the Kerr-Schild geometry, the metric and the electromagnetic
fields take different forms in different dimensions \cite{gur1},
\cite{gur2}. Here the only $D$ dependent quantity is the
cosmological constant. On the other hand, the energy density of
the null fluid depends on the acceleration parameter of the curve
$C$. It is worthwhile to look at the energy loss formula in this
case. Energy flux formula, in general, is given by (see
\cite{gur1}),

\begin{equation}
N=-\lim_{R \rightarrow \infty}\, \int_{S^{D-2}}\,
\dot{z}^{\alpha}\, T_{\alpha \beta}\, n^{\beta}\, R^{-1}\,
R_{0}^{D-2}\, d \Omega,
\end{equation}
where $T_{\mu \nu}$ is the corresponding energy momentum tensor
(of the fluid or the Maxwell field) and $n^{\mu}$ is a space-like
vector defined through $\lambda_{\mu}=-\dot{z}_{\mu}-\, n_{\mu}/R$
(see \cite{gur1} for more details). We find that $N_{F}=0$ for the
null fluid distribution. Its energy is conserved. The rate of
change of the energy of the electromagnetic field $N_{E}$ is found
as

\begin{equation}\label{ne}
N_{E}=\, e^2\, R_{0}^{D-4}\, \kappa_{1}^2\, \left\{-
\Omega_{D-3}\, \Gamma({D-2 \over 2})\, [-{\sqrt{\pi} \over
\Gamma({D-1 \over 2})}  +2^{D-1}\, {\Gamma({D+2 \over 2}) \over
\Gamma(D)}]+\Omega_{D-2} \right \},
\end{equation}
for $D \ge 4$, where $\Omega_{D}$ is the solid angle in $D$
dimensions. Here $\kappa_{1}$ is the first curvature of the curve
$C$. In calculating (\ref{ne}), we used the integration technique
developed in our earlier works \cite{gur1}. The quantity $N_{E}$
given above is positive for all dimensions $D$ and hence there is
energy loss due to acceleration in all dimensions. In particular,
when $D=4$, $N_{E}={8\pi \over 3}\,e^2\, \kappa_{1}^2$ is exactly
the energy loss formula due to the acceleration of a charged
particle in flat space-time \cite{jac}.

\vspace{0.3cm}

\vspace{0.3cm}

\noindent {\bf Remark 1}.\,We obtain the covariant derivative of
the Maxwell tensor $F_{\alpha \beta}$ as
\begin{equation}
\nabla_{\mu}\, F_{\alpha \beta}=\lambda_{\mu}\,
(\lambda_{\alpha}\,\zeta_{\beta}-\lambda_{\beta}\, \zeta_{\alpha})
\end{equation}
where
\[
\zeta^{\mu}={1 \over R}\, [{d^3\, {z}^{\mu} \over
d\tau_{0}^3}-a_{1}\, {d \,{z}^{\mu} \over d\tau_{0}}].
\]
This vector vanishes if $C$ is a straight line or has constant
acceleration, in which case $F_{\alpha \beta}$ becomes a
covariantly constant tensor field. Since $\lambda_{\mu}\,
\zeta^{\mu}=0$, then the current vector $J^{\mu}$ mentioned in
Theorem 2 is zero everywhere except on the curve $C$. In  fact it
takes the form

\[
J^{\mu}(x)={1 \over \Omega_{D-2}}\, \int_{C}\,
\dot{z}^{\mu}(\tau)\, \delta(x-z(\tau)) d\tau,
\]
where $\nabla_{\mu}\, J^{\mu}=0$ identically.

\vspace{0.3cm}

\noindent {\bf Remark 2}.\, We note that the Ricci scalar, the
Ricci invariant $R^{\alpha \beta}\,R_{\alpha \beta}$, and the
Maxwell invariant $F^{\alpha \beta}\, F_{\alpha \beta}$ are all
constants.

\begin{eqnarray}
g^{\alpha \beta}R_{\alpha \beta}&=&{(D-1)(D-4) \over
R_{0}^2},\\
R^{\alpha \beta}\,R_{\alpha \beta}&=&{ D^3-8D^2+21D-16 \over
R_{0}^4},\\
F^{\alpha \beta}\,F_{\alpha \beta}&=& {2 \over \kappa R_{0}^2}\,
(2-D).
\end{eqnarray}
Similarly the curvature invariant $R^{\alpha \beta \gamma
\sigma}\,R_{\alpha \beta \gamma \sigma}$ is also constant due to
the conformal flatness. All of these invariants are equal to the
corresponding invariants of the LBR metric (static case).

\section*{IV. Charged Particle with Constant Acceleration}

In this section we consider only the Einstein-Maxwell field
equations with a cosmological constant. To achieve this, we look
for special curves $C$ such that the null fluid energy density
$\rho$ vanishes.

Using the Serret-Frenet frame \cite{gur1} in $M_{D}$, we obtain
\[
\kappa \rho=(D-2)\,[-\dot{\kappa_{1}} \cos \theta +\kappa_{1}\,
\kappa_{2}\, \sin \theta \cos \phi],
\]
where $\kappa_{1}$ and $\kappa_{2}$ are the first two curvatures
of the curve $C$ and $(\theta, \phi)$ are the first two  angular
coordinates on $S^{D-2}$. It is clear that the energy density
$\rho$ of the null fluid does not have a fixed sign. It changes
its sign at different points. There is a non-trivial choice where
$\kappa_{1}$ is a non-zero constant and $\kappa_{2}=0$ which leads
to the vanishing of the energy density ($\rho=0$). In this case
$C$ describes a particle moving with a  constant acceleration and
the Maxwell field tensor $F_{\alpha \beta}$ is covariantly
constant. The total energy measured by the observer moving along
the curve $C$ is
\[
E=G_{\mu \nu}\,\dot{z}^{\mu}\, \dot{z}^{\nu}=(D-2)\,[\,
\kappa_{1}^2\, \sin^2 \theta +{D-3 \over 2R^2}\,].
\]
This vanishes asymptotically ($R \rightarrow \infty $) in the case
of LBR space-time, but asymptotically it is proportional  to the
square of the first curvature in the space-time corresponding to
our solution.

As an example of such a curve, let (using the notation
$x^{\mu}=(t,x,x^{2}, \cdots, x^{D-1})$)
\[
z^{\mu}=B \,(\sinh (w \tau_{0}),\, \cosh (w \tau_{0})-1,\, 0,\,
\cdots ,\, 0),
\]
where $B$ and $w$ are constants with $\kappa_{1}=w=1/B$, and
$\tau_{0}$ is defined through
\[
\cosh (w \tau_{0})={(x+B) [t^2-(x+B)^2-r^2-B^2] + 2B\,t\, R \over
2B[t^2-(x+B)^2]},
\]
where $r^2=(x^2)^2+(x^{3})^2+\cdots+(x^{D-1})^2$. Here R is the
retarded distance. It is given by
\[
R=\pm{1 \over
2B}\,\sqrt{(t^2-(x+B)^2-r^2-B^2)^2+4B^2\,(t^2-(x+B)^2)}.
\]
This curve has non-zero constant first curvature $\kappa_{1}$ and
all other curvatures $(\kappa_{i}=0,~~ i \ge 2)$ are zero. The
charged particle  has constant acceleration
$\sqrt{\ddot{z}^{\mu}\, \ddot{z}_{\mu}}=\kappa_{1}$ along the
$x$-direction.

In four dimensions since the cosmological constant is also zero,
we have a solution of the Einstein-Maxwell field equations
representing an (constant) accelerated charged particle. The LBR
metric and our solution given above are both conformally flat  and
solutions of the Einstein-Maxwell field equations. Furthermore, as
we mentioned in Remark 2, both have the same curvature invariants.
On the other hand LBR metric and our metric correspond to  two
distinct curves, namely, straight line and (non-zero) constant
curvature cases, respectively. In spite of this difference, these
two metrics are transformable to each other. Let

\begin{equation}\label{con}
{x^{\prime}}^{\mu}={x^{\mu}+sk^{\mu} \over 1 +2u +k^2\, s},
\end{equation}
where $s \equiv \eta_{\mu \nu}\, x^{\mu}x^{\nu}$, ~$u \equiv
k_{\mu}\,x^{\mu}$ and $k_{\mu}$ is a constant vector with $k^2
\equiv \eta_{\mu \nu} k^{\nu}k^{\mu}$. Here we choose $k_{0}=0$,
$Bk_{1}=-2$ and $k^{\mu}=0$ for all $\mu >1$. Then it is
straightforward to show that

\begin{equation}
ds^2={R_{0}^2 \over {R^{\prime}}^2}\, \eta_{\mu \nu}\,
d{x^{\prime}}^{\mu} d{x^{\prime}}^{\nu}={R_{0}^2 \over
x^2+r^2}\,\eta_{\mu \nu}\, dx^{\mu} dx^{\nu},
\end{equation}
where
\[
R^{\prime}=\pm{1 \over
2B}\,\sqrt{[(t^{\prime})^2-(x^{\prime}+B)^2-(r^{\prime})^2-B^2]^2+
4B^2\,(({t^{\prime}})^2-({x^{\prime}}+B)^2)}.
\]
This means that our solution is expressed in an (constant)
accelerated frame. This is the reason why we observe radiation.

\vspace{0.3cm}

\noindent {\bf Remark 3}.\, The infinitesimal version of the
conformal transformation (\ref{con}) (conformal Killing vector in
flat Minkowski space-time) is given in Penrose and Rindler
\cite{pr} as
\begin{equation}
\xi^{\mu}=-2ux^{\mu}+s k^{\mu}.
\end{equation}
They remark that `{\it ... the special conformal transformations
(four parameters sometimes misleadingly called uniform
acceleration transformation ...)}`. Here we observe that these
special transformations really correspond to a constant
acceleration.

\section*{V. Conclusion}

We found accelerated LBR metrics which solve Einstein-Maxwell-null
fluid field equations with a dimension dependent cosmological
constant. We have obtained the energy loss formula due to the
acceleration. In four dimensions it coincides with the standard
energy loss formula in flat space-time. We obtain the LBR metric
when the curve $C$ is a straight line in $M_{D}$. We also showed
that there is  another LBR limit. When the curve $C$ has constant
acceleration our solution is transformable (by a special conformal
mapping) to the LBR metric.

\vspace{1.5cm}

\section*{ Ackowledgment}
  \noindent This work is partially supported
by the Scientific and Technical Research Council of Turkey and by
the Turkish Academy of Sciences.



\begin{thebibliography}{99}

\bibitem{lev} T. Levi-Civita, {\it R.C. Acad. Lincei}, {\bf 26},
519 (1917).
\bibitem{ber} B. Bertotti, {\it Phys. Rev.}, {\bf 116}, 1331
(1959).
\bibitem{rob} R. Robinson, {\it Bull. Acad. Polon.}, {\bf 7}, 351 (1959).
\bibitem{kram} H. Stephani, D. Kramer, M. MacCallum, C.
Hoenselaers and E. Herlt, {\it Exact Solutions of Einstein's Field
Equations}, Cambridge Monographs on Mathematical Physics.
(Cambridge: Cambridge University Press). Second Edition (2003).
\bibitem{gib} G. W. Gibbons and C. M. Hull, {\it Phys. Lett}, {\bf
B109}, 190 (1982); G. W. Gibbons, in {\it Supersymmetry,
Supergravity and Related Topics}, eds. F. del Auglia, J. de
Azcaraga and L. Ibanez (World Scientific, Singapore, 1985). p.
147.
\bibitem{fer} S. Ferrara, {\it Bertotti-Robinson Geometry and
Supersymmetry}, Talk given at {\it 12th Italian Conference on
General Relativity and Gravitational Physics}, Rome, September,
1996\,\,({\tt hep-th/9701163}).
\bibitem{str} D. A. Lowe and A. Strominger, {\it Phys. Rev.
Lett.}, {\bf 73}, 1468 (1994).
\bibitem{kal} R. Kallosh and A. Peet, {\it Phys. Rev}, {\bf D46},
5223 (1992); R. Kallosh, {\it Phys. Lett}, {\bf B282}, 80 (1992).
\bibitem{gur4} M. G{\" u}rses, {\it Phys. Rev.}, {\bf D46}, 2522
(1992).
\bibitem{cor} V. Cardoso, O. J. C. Dias, J. P. S. Lemos, {\it
Phys. Rev.}, {\bf D70}, 024002 (2004).
\bibitem{gur5} M. G{\" u}rses and E. Sermutlu, {\it Class. Quantum
Grav.}, {\bf 12}, 2799 (1995).
\bibitem{bv1} W. B. Bonnor and P. C. Vaidya, {\it General Relativity},
papers in honor of J. L. Synge, edited by L. O' Raifeartaigh
(Dublin: Dublin Institute for Advanced Studies) p. 119 (1972). \
\bibitem{gur1} M. G{\" u}rses and {\" O}. Sar{\i}o\u{g}lu, {\it Class.
Quantum Grav.}, {\bf 19}, 4249 (2002); Erratum, {\it ibid}, {\bf
20}, 1413 (2003).
\bibitem{gur2} M. G{\" u}rses and {\" O}. Sar{\i}o\u{g}lu, {\it Class.
Quantum Grav.}, {\bf 20}, 351 (2003).
\bibitem{ozg} {\" O}. Sar{\i}o\u{g}lu, {\it Phys. Rev.}, {\bf D66}, 085005 (2002).
\bibitem{gur3} M. G{\" u}rses and {\" O}. Sar{\i}o\u{g}lu, {\it J.
Math. Phys.}, {\bf 44}, 4672 (2003).
\bibitem{tan} F. R. Tangherlini, {\it Nuovo Cimento}, {\bf 77}, 636 (1963).
\bibitem{pleb} A. Garc{\' {\i}}a, I. H. Salazar and J. F. Pleba{\' n}ski, {\it
Nuovo Cimento}, {\bf B84}, 65 (1984).
\bibitem{trt} A. Trautman, {\it Phys. Rev. Lett.}, {\bf 46}, 875 (1981).
\bibitem{jac} J. D. Jackson, {\it Classical Electrodynamics}, (New
York: Wiley) (1975).
\bibitem{pr} R. Penrose and W. Rindler, {\it Spinors and
Space-Time: Spinor and twistor methods in space-time geometry},
Vol.2 (Cambridge: Cambridge University Press). p 83-84, (1986).
\end{thebibliography}
\end{document}